%% file: small-R-june.tex
\documentclass{article}

\usepackage{graphicx}
\usepackage{feynmf}

\input epsf
\def\beq{\begin{equation}}   \def\eeq{\end{equation}}
\def\bea{\begin{eqnarray}}  \def\eea{\end{eqnarray}} \def\nn{\nonumber}
\def\noi{\noindent} \def\beeq{\begin{eqnarray}}
\def\eeeq{\end{eqnarray}}
\def\lsim{\raise0.3ex\hbox{$<$\kern-0.75em\raise-1.1ex\hbox{$\sim$}}}
\def\gsim{\raise0.3ex\hbox{$>$\kern-0.75em\raise-1.1ex\hbox{$\sim$}}}

\newcommand\mysection{\setcounter{equation}{0}\section}

\newcounter{hran}

\begin{document}

\begin{fmffile}{samplepics}
\setlength{\unitlength}{1mm}

\begin{titlepage}

\vspace{1.cm}

\begin{center}

{\large \bf ISOLATING PROMPT PHOTONS WITH NARROW CONES}\\[2cm]

{\large S. Catani$^{a}$, M. Fontannaz$^{b}$, J.~Ph.~Guillet$^{c}$ and
E.~Pilon$^{c}$} \\[.5cm]

\normalsize
{$^{a}$ INFN, Sezione di Firenze and Dipartimento di Fisica e Astronomia,\\ 
Universit\`a di Firenze, I-50019 Sesto Fiorentino, 
Florence, Italy}
\\[.2cm]
{$^{b}$ Laboratoire de Physique Th\'eorique LPT,}\\
{UMR 8627 du CNRS, Universit\'e de Paris XI, F-91405 Orsay Cedex, France}
\\[.2cm]        
{$^{c}$ LAPTh, Univ. de Savoie, CNRS, B.P. 110, Annecy-Le-Vieux, F-74941, 
France}\\
      
\end{center}

\vspace{4cm}

\begin{abstract} 
\noindent
We discuss the isolation of prompt photons in hadronic collisions by means of 
narrow isolation cones and the QCD computation of the corresponding cross 
sections. We reconsider the occurence of large perturbative terms with 
logarithmic dependence on the cone size and their impact on the fragmentation
scale dependence. We cure the apparent perturbative violation of unitarity for 
small cone sizes, which had been noticed earlier in next-to-leading-order (NLO)
calculations, by resumming the leading logarithmic dependence on the cone size.
We discuss possible implications regarding the implementation of some hollow
cone variants of the cone criterion, which simulate the experimental difficulty
to impose isolation inside the region filled by the electromagnetic shower that
develops in the calorimeter.
\end{abstract}

\vspace{3cm}


\vspace{2cm}

\end{titlepage}

\pagestyle{plain} 

\input{introduction-exp}

\input{art3-apr}

\input{hollow-short}

\input{bibliography}

\end{fmffile}

\end{document}

%% file: introduction-exp.tex
\mysection{Introduction}

The measurements of hard prompt photons by TeV collider experiments, such as
those that were performed at the Fermilab Tevatron 
\cite{Aaltonen:2009ty,Abazov:2005wc},
RHIC \cite{Adare:2012yt}, and those currently carried out at the CERN LHC 
\cite{Aad:2011tw,Chatrchyan:2011ue}, 
have been long recognised as both important
tests of QCD and  Standard Model (SM) benchmarks useful to back up the
understanding of the LHC  detectors at the begining of the LHC era. 
Moreover, photon production and, especially,  photon pair 
production\footnote{References \cite{Aaltonen:2009ty} to 
\cite{Chatrchyan:2011qt} 
include some of the most recent experimental results on isolated photons in 
hadron collisions and further references to previous experimental results.}
\cite{Aaltonen:2012jd,Abazov:2010ah,Aad:2011mh,Chatrchyan:2011qt} 
provide large SM backgrounds to signatures of various potential effects
of new  physics 
\cite{Aaltonen:2009in,Abazov:2008kp,Aad:2012fw,Chatrchyan:2012tea}. 
In this respect, one should distinguish the so-called
{\em prompt} photons from what could be named {\em secondary} photons. Prompt
photons directly take part in the hard partonic subprocess. Secondary photons
instead originate from the decays of hadrons (e.g., $\pi^{0}$ and $\eta$
mesons) that are produced in the final (subsequent to the hard-scattering
process) parton-to-hadron fragmentation stage of the hadronic collision. The
huge yield of secondary photons at colliders overwhelms the production of prompt
photons. This is even more true for signals of new physics involving photons
with moderately large transverse momenta ($p_{T}$), such as, for example, the
search for the Higgs boson at LHC in the two-photon decay channel in the mass
range 120-130~GeV 
\cite{ATLAS:2012ad,Chatrchyan:2012twa}. 

\vspace{0.3cm}

\noindent
To reduce the background of secondary photons, collider experiments impose
isolation cuts. The isolation cuts, which are optimized to the experimental
setup, act on calorimetric deposits, tracks, discriminating shape observables
and so forth, in a quite sophisticated way. Such criteria can be taken into
account in full-fledged event simulations but not in calculations performed at
the parton level only. 
A customary basic requirement that can be implemented in parton level
calculations is transverse-energy isolation.
One considers the photon candidate and the direction of its momentum $p^\gamma$,
as specified by the rapidity and azimuthal-angle variables $y_\gamma$ and
$\phi_\gamma$ (rapidities, azimuthal angles and transverse energies are defined
in a reference frame where the momenta of the two colliding hadrons are
back-to-back). Around the direction of the photon candidate, one draws a cone
${\cal C}_{\gamma}(R)$ of aperture $R$ in $\{ y, \phi\}$ space and considers the
hadrons (each hadron with momentum $p_k$ and corresponding transverse energy
$E_{T}^{(k)}$) inside the cone:
\begin{equation}\label{e1-1}
\mbox{hadron $k$ $\in$} \; {\cal C}_{\gamma}(R) 
\;  \Leftrightarrow  \;
\sqrt{ \left( y_{k} - y_{\gamma} \right) ^{2} \, + \, 
       \left( \phi_{k} - \phi_{\gamma} \right) ^{2} } \leq R \;\;.
\end{equation}
Then one requires that the total amount of hadronic 
transverse energy inside this cone is smaller than a maximum amount 
$E_{T}^{iso}$:
\begin{equation}\label{e1-2}
\sum_{k \,\in \,{\cal C}_{\gamma}(R)} E_{T}^{(k)} \, 
< E_{T}^{iso} \;\;.
\end{equation}
Recent improvements of this isolation criterion recommend to first subtract the
contribution coming from the underlying event and from pile-up, before the
criterion is  applied; the contribution to be subtracted is assessed from the
mini-jet activity away from the direction of the photon candidate
\cite{sub-underlying}. Other criteria at variance with that of
eqns.~(\ref{e1-1}) and (\ref{e1-2})  have also been proposed 
\cite{democratric1,democratric2,frixione,frix-discretise}. 
All these isolation criteria have impact on the selected sample of prompt
photons. As described for instance in refs.~\cite{1r, diphox}, prompt photons
may be schematically viewed as produced by two  mechanisms: the ``direct" (D)
mechanism, in which the photon is produced directly at high $p_{T}$ by hard
scattering, and the ``fragmentation" (F) mechanism, in which the photon
originates from the (essentially collinear) fragmentation of a high-$p_{T}$
coloured parton primarily produced by hard scattering. The isolation criteria
have impact on both ``direct" and ``fragmentation" photons. In particular, the
production rate through the fragmentation mechanism is strongly reduced by the
isolation, since  the ``fragmentation" photon is generally produced inside a
large-$p_{T}$ jet of hadrons (unless the photon carries a major fraction of its
parent parton's transverse momentum).

\vspace{0.3cm}

\noindent
QCD radiative corrections for isolated prompt-photon production at hadron
colliders have been computed in the literature.
The next-to-leading order (NLO) QCD corrections to single-inclusive photon
production were computed in ref.~\cite{Gordon:1993qc} (using cone isolation in
the small-$R$ approximation) and ref.~\cite{1r} (for any infrared-safe isolation
criteria). Diphoton production has been computed at the NLO
\cite{diphox, Campbell:2011bn}, including NLO corrections \cite{Bern:2002jx}
to the gluon fusion channel, and at the next-to-next-to-leading order (NNLO)
\cite{Catani:2011qz} (using the isolation criterion of ref.~\cite{frixione}).
The NLO calculation of `photon plus one jet' was performed 
in ref.~\cite{Belghobsi:2009hx}. Diphoton production in association with 
one jet \cite{DelDuca:2003uz} and two jets \cite{Bern:2011pa} has been computed
at the NLO by using the isolation criterion of ref.~\cite{frixione}.
The NLO calculation of `diphoton plus one jet' for general isolation criteria
has been performed recently \cite{Gehrmann:2013aga}.
Higher-order QCD contributions due to soft gluons 
\cite{Balazs:2007hr, Becher:2009th}, 
high-energy logarithmic corrections \cite{Baranov:2007np, Diana:2010ef},
and parton shower effects \cite{Hoeche:2009xc, D'Errico:2011sd, Odaka:2012ry}
have also been studied.

\vspace{0.3cm}

\noindent
In the present article we will focus on the ``standard cone criterion" defined by
eqns.~(\ref{e1-1}) and(\ref{e1-2}), and on its implementation in QCD calculations 
at partonic level.
We will also discuss some implications for the implementation of a  `two cone'
criterion that aims at simulating a poorer isolation around the electromagnetic
cluster of  photon candidates in some experimental configurations. In
ref.~\cite{1r} we studied the standard cone criterion, and we presented the
calculation of isolated-photon cross sections at the NLO of the perturbative
expansion in powers of the QCD coupling $\alpha_{s}$. In particular, we studied
the dependence of the cross section on the size $R$ of the isolation cone.
Considering small values of $R$ (typically $R \,\lsim \,0.1$), we noticed
\cite{1r} a violation of unitarity of the NLO result, since the NLO isolated
cross section becomes {\em larger} than the NLO inclusive (i.e., without
isolation) cross section. Therefore, at small values of $R$, the NLO result is
certainly unphysical. Moreover, this finding shed doubts on the reliability of
the NLO QCD prediction also at
moderately-small values of $R$ ($R \sim 0.4$-0.3) that are actually used in
experiments.
The purpose of this article is to trace back this misbehaviour of the NLO result
and to cure it. 

\vspace{0.3cm}

\noindent
The paper is organized as follows.
We detail how the $R$ dependence, which is dominantly logarithmic at small 
$R$ order by order in perturbation theory, appears in both the NLO calculation
(sec.~\ref{nlo-calc-R}) and at higher-order levels (sec.~\ref{secallorder}).
We point out how the isolation constraint on transverse energy causes 
a mismatch in the $R$ dependence produced by parton radiation inside and outside
the isolation cone. This mismatch produces the observed violation of unitarity 
in the NLO calculation at small values of $R$, and it makes
an all-order resummation of the ensuing $\ln R$ terms mandatory.
In sec.~\ref{isol} we discuss the implementation of resummation to 
leading logarithmic (LL) accuracy for the standard cone isolation criterion.
Then, in sec.~\ref{results}, we present numerical results at Tevatron and LHC 
energies, and we explicitly show how LL resummation removes the unphysical
behaviour of the NLO calculation at small values of $R$.
In sec.~\ref{impli-hgc} we discuss  some implications of these results for the
implementation of a criterion based on a hollow  cone,  which (very crudely) mimics
the difficulty to experimentally implement isolation in the solid angle  filled by
the electromagnetic cluster of a hard photon in a detector.  A brief summary is
presented in sec.~\ref{secsum}.

%% file: art3-apr.tex

\mysection{Origin of the logarithmic dependence on the cone size at the
NLO}\label{nlo-calc-R}

We consider the isolation criterion in eqns.~(\ref{e1-1}) and (\ref{e1-2}) and
the ensuing small-$R$ behaviour of the isolated-photon cross section at the NLO.
To identify the origin of the logarithmic dependence on $R$, we briefly recall
the results obtained in refs.~\cite{Gordon:1993qc, 1r} on the calculation of the
higher-order (HO) contribution to the Born level cross section.

\vspace{0.3cm}

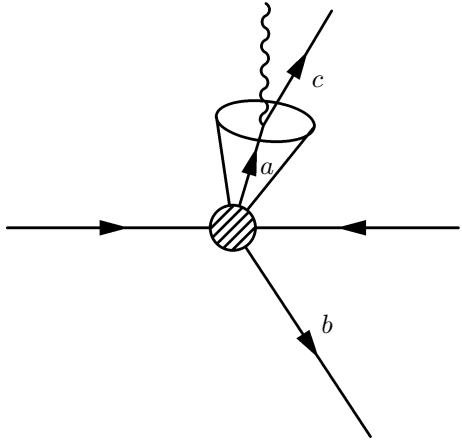
\begin{figure}[h]
\centering
\[
\parbox[c][50mm][c]{50mm}{\begin{fmfgraph*}(60,60)
  \fmfleft{q1}
  \fmfright{q2}
  \fmftop{t1,t2,t3,t6,ph,q,t7,t8}
  \fmfbottom{b1,b2,b3,b4,b,b6}
  \fmf{fermion}{q1,v}
  \fmf{fermion}{q2,v}
  \fmffreeze
  \fmf{fermion,tension=2.3,label=$a$,label.dist=0.02w}{v,v1}
  \fmf{fermion,label=$b$}{v,b}
  \fmf{fermion,label=$c$}{v1,q}
  \fmf{photon}{v1,ph}
  \fmfblob{.1w}{v}
  \fmf{plain}{v,v6}
  \fmf{phantom}{v6,t6}
  \fmf{plain}{v,v7}
  \fmf{phantom}{v7,t7}
  \fmffreeze
  \fmfposition
  \fmfipair{p[]}
  \fmfiset{p6}{vloc(__v6)}
  \fmfiset{p7}{vloc(__v7)}
  \fmfiset{p0}{vloc(__v)}
  \fmfiequ{x0}{xpart(p0)}
  \fmfiequ{y0}{ypart(p0)}
  \fmfiequ{x6}{xpart(p6)}
  \fmfiequ{y6}{ypart(p6)}
  \fmfiequ{x7}{xpart(p7)}
  \fmfiequ{y7}{ypart(p7)}
  \fmfiequ{x1}{(x6+x7)/2}
  \fmfiequ{y1}{(y6+y7)/2}
  \fmfiset{p1}{(x1,y1)}
  \fmfiequ{a}{(y6-y7)/(x6-x7)}
  \fmfiequ{al}{angle(1,a)}
  \fmfiequ{t}{(x1-x6)++(y1-y6)}
  \fmfiset{t}{2*t}
  \fmfpen{thin}
  \fmfcmd{draw fullcircle xscaled t yscaled 15pt rotated al shifted (x1,y1);}
\end{fmfgraph*}}
\]
\caption{A pictorial representation of the NLO contribution from final-state 
collinear radiation inside the isolation cone}
\label{fig1} 
\end{figure}

\noindent
We start with the contribution where the photon is accompanied by
collinear parton radiation inside a 
cone of radius $R$ (see fig.~\ref{fig1}). This
part of the HO correction to the Born level direct (D)
cross section leads to the NLO fragmentation (F)
contribution.
Using dimensional regularization in $d=4-2\epsilon$ space-time dimensions,
the differential cross section with respect to the transverse momentum 
$\vec{p}_{T}^{\, \gamma}$ and the rapidity (or, equivalently,
pseudorapidity) $\eta ^{\gamma}$ of the photon is
(we refer the reader to ref.~\cite{1r} for more details about the notation)
\bea
\lefteqn{
\left. 
 \frac{d \sigma}{d \vec{p}_{T}^{\, \gamma} \, d\eta ^{\gamma}} 
\right| 
^{(HO)}_{coll \, inside \, cone}
}
\nonumber\\ 
& = & \!\!\! \left ( {\alpha_s \over \pi} \right )^2 
\int_{z_{min}}^1 
{d\sigma^{Born} \over d\vec{p}_{T}^{a} \, d\eta^a \, d\eta^b} (A + B \to a + b) \,
d\eta^b \, {dz \over z^2} \, e_a^2 \, {\alpha \over 2 \pi} 
\nonumber\\
&& \!\!\!
\left\{ 
P_{q \gamma}(z)
 \left[ 
  - \frac{(4 \pi)^\epsilon}{\epsilon} 
    \frac{\Gamma ( 1 - \epsilon )}{\Gamma (1 - 2 \epsilon )} 
  + \ln \left( \frac{M_F^2}{\mu_{reg}^2} \right) 
  + \ln \left( \frac{R^{2} (p_{T}^{\gamma})^{2}}{M_{F}^{2}} \right)
  + \ln \left( (1-z)^2  \right)
 \right] 
 + z  
\right\} \,.
\label{2.1e}
\eea

\noi 
The factor
\[
{d\sigma^{Born} \over d\vec{p}_{T}^{a} \, d\eta^{a} \, d\eta^{b}} (A + B \to a + b)
\]
is the Born level cross section 
(the overall power of $\alpha_s$ is not included in $d\sigma^{Born}$ and it is
explicitly denoted in eqn.~(\ref{2.1e}))
of the reaction hadron $A$ + hadron $B \to$
parton $a$ + parton $b$. In the NLO contribution of eqn.~(\ref{2.1e}), the
fragmenting parton $a$ and the collinear parton $c$ are either quarks or
antiquarks.

\vspace{0.3cm}

\vspace{0.3cm}

\noi
To obtain the expression on the right-hand side of
eqn.~(\ref{2.1e}), the integration over the angular phase space of parton $c$ 
has been restricted to lie inside the cone and, moreover,
we have used the small-cone approximation, thus neglecting terms of ${\cal O} (R^2)$. 
The integration over the momentum fraction $z=p_{T}^{\gamma}/p_{T}^{a}$
is bounded by $z_{min}$, which is fixed by the photon kinematics and the
centre--of--mass energy $\sqrt S$ of the two colliding hadrons.
The dimensional regularization scale is denoted by $\mu_{reg}$, and for later
convenience
we have introduced
the auxiliary factorization scale $M_F$ (the right-hand side of
eqn.~(\ref{2.1e}) is actually independent of $M_F$).

\vspace{0.3cm}

\noindent
The first two terms in the curly bracket of eqn.~(\ref{2.1e}) correspond 
to (the $\epsilon$-expansion of) the bare
photon fragmentation function $D_a^{\gamma \, bare}$ in the customary
${\overline {\rm MS}}$ factorization scheme,
\beq
\label{2.2e}
D_a^{\gamma \, bare} (z, M_F, \epsilon ) 
= 
- {1 \over \epsilon} {\Gamma (1 - \epsilon ) \over \Gamma (1 - 2 \epsilon )} 
\left ( {4 \pi \mu_{reg}^2 \over M_F^2}\right )^\epsilon \ K_{a}^{(0)} (z) \;\;,
\eeq

\noi 
with 
\begin{equation}\label{Kq0}
K_{a}^{(0)} (z) 
=  
\frac{\alpha}{2 \pi} \ e_a^2 \ \frac{1 + (1 - z)^2}{z}
\equiv 
\frac{\alpha}{2 \pi} \ e_a^2 \ P_{q \gamma}(z) 
\end{equation}
and where $\alpha$ is the fine structure constant and $e_a$ is the electric charge
of the parton $a$ ($K_{g}^{(0)} (z)=0$ since $e_g=0$).
Within the ${\overline {\rm
MS}}$ factorization scheme,
the other terms 
in the curly bracket of eqn.~(\ref{2.1e}) are considered
as HO corrections to the direct cross section coming 
from parton radiation inside the cone. The proper treatment of the collinear $1/\epsilon$ singularity to all 
orders (as discussed in details, e.g., in refs.~\cite{2r,3e}) leads to the introduction of 
the all-order fragmentation functions $D_a^{\gamma}(z, M_F)$. These fragmentation functions obey the 
following inhomogeneous evolution equations: 
\beq
\label{2.3e}
M_F^2 {\partial D_a^{\gamma} \over \partial M_F^2} 
= K_a + \sum_b P_{ba} \otimes D_b^{\gamma} \;\;,
\eeq

\noi 
where the symbol $\otimes$ denotes the following convolution: 
\[
(f \otimes g) (z) =  \int_0^1 du \, \int_0^1 \, dv \, f(u) \, g(v) \, \delta (uv -z) \;.
\] 
The all-order functions $K_a(z)$ are the inhomogeneous kernels for the
collinear splitting 
`parton $a$ to photon',
\[
K_a(z) = K_a^{(0)}(z) + {\alpha_s \over 2 \pi} K_a^{(1)}(z) + \cdots \;,
\] 
where the leading order (LO) term $K_a^{(0)}(z)$ 
is given in eqn.~(\ref{Kq0}). The all-order functions
$P_{ab}(z)$ are the usual Dokshitzer--Gribov--Lipatov--Altarelli--Parisi 
(DGLAP) time-like splitting kernels,
\[
P_{ba}(z) 
= 
\frac{\alpha_{s}}{2 \pi} \ P^{(0)}_{ba}(z) 
+ \left( \frac{\alpha_{s}}{2 \pi} \right) ^{2} \ P_{ba}^{(1)}(z) 
+ \cdots \;\;,
\]
where $P^{(0)}_{ba}(z)$ is the LO term, $P^{(1)}_{ba}(z)$ is the NLO
term and so forth.
A thorough discussion of the evolution equation (\ref{2.3e}) and its solutions 
with appropriate boundary conditions
can be found in ref.~\cite{2r}.

\vspace{0.3cm}

\noindent
The contribution coming from the integration over the phase space region where parton $c$ 
is outside the cone contains a term proportional to $\ln(1/R)$. When no isolation is imposed 
on the collinear 
debris that accompanies the photon,
the $\ln (1/R)$ dependence from outside the cone completely cancels against the
$\ln R$ term of eqn.~(\ref{2.1e}). 
On the contrary, in the case of the isolated cross section, when the parton $c$ lies inside the cone 
the isolation requirement of eqn.~(\ref{e1-2}) leads to the constraint
$p_{T}^{c} \leq E_{T}^{iso}$
($p_{T}^{c}=(1-z) p_{T}^{a}= (1-z) p_{T}^{\gamma}/z$), 
which restricts the integration range over $z$ 
in eqn.~(\ref{2.1e}) to the following region:
\beq
\label{2.4e}
z \geq z_{cut} \equiv \frac{1}{1+\varepsilon} \;\;\;\;\;, \;\;\;\;\; 
\varepsilon \equiv \frac{E_{T}^{iso}}{p_{T}^{\gamma}}
\eeq

\noi 
This restriction produces a
mismatch of the $\ln R$ dependences from inside vs. outside the cone, and this
leads to a net $\ln R$
dependence in the isolated cross section. At the NLO
this $\ln R$ dependence is given by the term
\beq
\label{2.5e}
- \left ( {\alpha_s \over \pi} \right )^2 \int_{z_{min}}^{z_{cut}} 
{d\sigma^{Born} \over d\vec{p}_{T}^{a} d \eta^a d\eta^b} d\eta^b {dz \over z^2} 
K_{a}^{(0)} (z) \, \ln (R^2)
\eeq

\noi that blows up towards $+ \infty$ when $R \to 0$. This unbounded increase
of the NLO isolated cross section leads to the violation of unitarity that 
was observed in ref.~\cite{1r}. As pointed out in ref.~\cite{1r}, this
unphysical effect
is an artefact of the fixed-order
truncation of the QCD perturbative series: an all-order summation in $\alpha_s$
of the $\ln R$ 
terms should cure the pathological behaviour induced at the NLO
by the contribution in eqn.~(\ref{2.5e}).

\vspace{0.3cm}

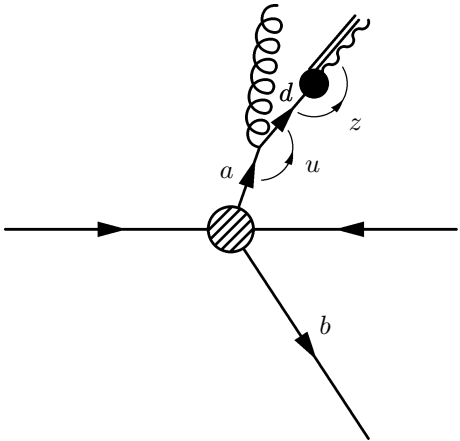
\begin{figure}[h]
\centering
\[
\parbox[c][50mm][c]{50mm}{\begin{fmfgraph*}(60,60)
  \fmfleft{q1}
  \fmfright{q2}
  \fmftop{t1,t2,t3,g,ph,t6}
  \fmfbottom{b1,b2,b3,b4,b,b6}
  \fmf{fermion}{q1,v}
  \fmf{fermion}{q2,v}
  \fmffreeze
  \fmf{fermion,tension=2.5,label=$a$}{v,v1}
  \fmf{fermion,label=$b$}{v,b}
  \fmf{fermion}{v1,v2}
  \fmf{gluon}{g,v1}
  \fmf{photon}{v2,ph}
  \fmfblob{.1w}{v}
  \fmfv{decor.shape=circle,decor.filled=full,decor.size=.06w,label=$d$,label.angle=200,label.dist=.05w}{v2}
  \fmffreeze
  \fmfposition
  \fmfi{plain}{vpath (__v2,__ph) shifted (thick*(-2.5,0.8))}
  \fmfi{plain}{vpath (__v2,__ph) shifted (thick*(-1.8,0))}
  \fmfdraw
  \fmfipath{c[]}
  \fmfipair{p[]}
  \fmfiset{p1}{vloc(__v1)}
  \fmfiset{p2}{vloc(__v2)}
  \fmfset{thin}{0.5pt}
  \fmfset{arrow_len}{2mm}
  \fmfpen{thin}
  \fmfiset{c1}{fullcircle scaled 25pt shifted p1}
  \fmfiset{c2}{fullcircle scaled 25pt shifted p2}
  \fmfi{fermion,label=$u$}{ subpath (-length(c1)*0.24,length(c1)/12) of c1}
  \fmfi{fermion,label=$z$}{ subpath (-length(c2)*0.33,length(c2)/12) of c2}
\end{fmfgraph*}}
\]
\caption{A HO correction to the fragmentation contribution}
\label{fig2}
\end{figure}

\noindent
We have so far discussed the HO correction to the Born level direct cross
section. An analogous discussion applies to the HO
correction to the Born level fragmentation component of the cross section
(fig.~\ref{fig2}).
In this case the photon production process proceeds
through the fragmentation function $D^\gamma_d$ (which contains a perturbative 
and a non-perturbative component) of a parton $d$.
The effects of collinear parton radiation inside and outside the cone around
the parton $d$, and the ensuing mismatch of the $\ln R$ dependence in the NLO
isolated cross section produce a contribution
that is analogous to the term in eqn.~(\ref{2.5e}).
The main difference with respect to eqn.~(\ref{2.5e}) is that the factor
$K_{a}^{(0)} (z) \, \ln (R^2)$ is replaced by the factor
\beq
{\alpha_s \over 2\pi} \; P^{(0)}_{da}(z/x) \, \ln (R^2)
\eeq
that is then convoluted with the fragmentation function $D_d^{\gamma}(x, M_F)$.
At fixed values of $M_F$, this $\ln R$ contribution to the fragmentation
component of the NLO isolated cross section also 
blows up towards $+ \infty$ when $R \to 0$. Therefore, the violation
of unitarity produced by the NLO direct contribution is not removed by 
the NLO fragmentation contribution.
The all-order summation 
of the $\ln R$ 
terms should cure the pathological behaviour observed at the NLO
in both the direct and fragmentation components of the isolated 
cross section.\par

\mysection{Multiparton collinear radiation and $\ln R$ 
dependence}\label{secallorder}

The collinear-radiation spectrum of a {\em single} parton that is emitted
inside a small cone and just outside of it
produces a $\ln R$ contribution to the NLO cross section.
This effect has been briefly recalled in sec.~\ref{nlo-calc-R}. In the present
section, we consider the effect of {\em multiple} collinear radiation,
and we illustrate how the $\ln R$ contributions
arise at all orders in QCD perturbation theory. The discussion of this section
sets the stage for the resummation of the $\ln R$ dependence of the isolated cross
section, which is considered in sec.~\ref{isol}.
 
\vspace{0.3cm}

\noi The discussion is more conveniently formulated by taking Mellin moments of the relevant 
expressions. In the following we use a shorthand notation.
For instance, the $n$-moment $\sigma (n)$  of the differential cross section 
${d\sigma \over dp_{T}^{\gamma} d\eta^{\gamma}}$ is defined as follows:
\beq
\label{2.6e}
\sigma (n) 
= 
\int_0^1 dx_{T}^{\gamma} \ \left( x_{T}^{\gamma} \right) ^{n-1} \ {d\sigma \over dp_{T}^{\gamma} 
d\eta^{\gamma}}\ , \;\;\;\; x_{T}^{\gamma} \equiv {2p_{T}^{\gamma} \over \sqrt{S}}
\;,
\eeq
and eqn.~(\ref{2.1e})
\noi can be written as
\beq
\label{2.7e}
\sigma^{(HO)} (n) 
= 
\left( {\alpha_s \over \pi} \right) ^2 \sigma_a^{Born}(n) 
\left[ 
 D_a^{\gamma \, bare} (n, M_F, \epsilon ) 
 +
 K_{a}^{(0)} (n) \, \ln \left( {R^2 (p_{T}^{\gamma})^2 \over M_F^2} \right)
 + 
 t_{a}(n) 
\right] \;,
\eeq
with 
\bea
D_a^{\gamma \, bare}(n,M_F, \varepsilon ) 
& = & \int_0^1 dz\ z^{n-1} 
D_a^{\gamma \, bare} (z, M_F, \epsilon ) 
\nonumber\\
& = & 
- {1 \over \epsilon} {\Gamma (1 - \epsilon ) \over \Gamma (1 - 2 \epsilon )} 
\left ( {4 \pi \mu_{reg}^2 \over M_F^2}\right )^\epsilon \ K_{a}^{(0)} (n) \;,
\label{2.8e}
\eea
and where we have introduced
\begin{equation}\label{Kn}
K_{a}^{(0)} (n) = {\alpha \over 2 \pi} e_a^2 
\int_0^1 dz\ z^{n-1} \ P_{q \gamma}(z) \;,
\end{equation}
\begin{equation}\label{tn}
t_a(n) = {\alpha \over 2 \pi} e_a^2 
\int_0^1 dz\ z^{n-1} \ 
\left[ \frac{1 + (1-z)^2}{z} \ln \left( (1 - z)^2 \right) + z \right] \;.
\end{equation}

\vspace{0.3cm}
\noi
The all-order evolution equation (\ref{2.3e}) written in Mellin moments leads to a solution in closed 
form. We write it here explicitly for the flavour non-singlet (NS)
component\footnote{In the NS 
component
the factor $e_a^2$ in $K_q^{(0)}$ has to be replaced by 
$2 ( e_a^2 - \left< e_a^2 \right> )$, where
$\left< e_a^2 \right> = \frac{1}{N_f} \, \sum_{b=1}^{N_f} e_b^2$.
} (and we drop all flavour indices, for the sake of simplicity):
\beq
\label{2.9e}
D(n, M_F) 
= 
\int_{M_0^2}^{M_F^2} {dk^2 \over k^2} K(n) \ 
e^{\int_{k^2}^{M_F^2} {dk{'}^2 \over k{'}^2} P(n)} 
+ \; e^{\int_{M_0^2}^{M_F^2} {dk{'}^2 \over k{'}^2} P(n)} \;D(n, M_0)\;,
\eeq
\noi 
where the splitting kernels $K(n)$ and $P(n)$ are those of the NS case.
Considering the first term in the right-hand side of eqn.~(\ref{2.9e}) and
taking the lower bound of the $k^2$-integral equal to zero leads to collinear 
divergences, whose dimensional regularization produces $1/\epsilon$ poles  
as in eqns.~(\ref{2.2e}) and (\ref{2.8e}). 
Following the 
procedure described
in refs.~\cite{2r,3e},
in the expression (\ref{2.9e})
dimensional regularization has been replaced by the cut-off $M_0$, which
corresponds to the boundary between the perturbative and non-perturbative
domains. 
Expanding the first term  in the right-hand side of eqn.~(\ref{2.9e}), we indeed recover the
perturbative expression of zeroth order in $\alpha_{s}$ of the NS fragmentation function, 
\begin{equation}\label{ns-eq}
D^{(LO)}(n, M_F) 
= 
2 \left( e_a^2 - \left< e_a^2 \right> \right) {\alpha \over 2 \pi }
P_{\gamma q}(n) \ln \left( \frac{M_F^2}{M_0^2} \right) \;\;.
\end{equation}
The second term in the right-hand side of eqn.~(\ref{2.9e}) is thus
proportional to $D(n, M_0)$, which is the non-perturbative contribution
to the fragmentation function at the cut-off scale $M_0$.

\vspace{0.3cm}

\noindent
The all-order perturbative structure of the cross section in eqn.~(\ref{2.6e})
follows from the QCD factorization theorem of collinear singularities. Considering
the NS contribution to the cross section,
the factorization theorem allows us to write 
\beq
\label{2.10e-new}
\sigma(n) = \sigma_{F}(n,M_F) \; D(n, M_F) + \sigma_{D}(n,M_F) \;\;,
\eeq
where the first term on the right-hand side is the fragmentation component,
and the second term is the direct component (this separation in two components
depends on the factorization scale $M_F$).
The partonic subprocess cross sections $\sigma_{F} (n,M_F)$
and $\sigma_{D}(n,M_F)$ have an expansion in $\alpha_s$ of the form
\beq
\label{2.11e}
\sigma_{F}(n,M_F) = \left ( {\alpha_s \over \pi} \right )^2
\sigma^{Born}_{F}(n) +  \sigma^{(1)}_{F}(n,M_F) + \cdots \;,
\eeq
\beq
\label{2.11edir}
\sigma_{D}(n,M_F) =  {\alpha_s \over \pi} \;\;
\sigma^{Born}_{D}(n) +  \sigma^{(1)}_{D}(n,M_F) + \cdots \;,
\eeq
with $\sigma^{(k)}_{F} \sim {\cal O} (\alpha_s^{2+k})$
and $\sigma^{(k)}_{D} \sim {\cal O} (\alpha_s^{1+k})$.


\vspace{0.3cm}

\noindent
We are interested in computing and resumming the $\ln R$-enhanced perturbative
terms that arise in the presence of an isolation cone of small size $R$. Before
studying the effect of isolation,
we first consider the {\em non-isolated} cross section to all orders.
We partition the available phase space by introducing a cone of size $R$ around
the photon, and we consider the effect of QCD radiation inside and outside this
cone. This partition artificially splits 
the cross section into terms that separately and explicitly depend on $R$,
although the 
complete cross section is independent of $R$.
To compute the all-order $\ln R$-dependence produced by this splitting,
we exploit the basic physical picture that underlies the factorization structure
of eqn.~(\ref{2.10e-new}). Indeed, the photon fragmentation function originates
from the resummation of multiparton collinear radiation that is produced around
the photon direction.
Roughly speaking, the fragmentation function $D(n, M_F)$ embodies parton
radiation inside a cone whose radius is of the order of $M_F/p_{T}^{\, \gamma}$.
Moreover, the all-order evolution equation
(\ref{2.3e}) follows from the angular ordered
structure of multiple QCD radiation around the photon direction. Thus, the
solution in eqn.~(\ref{2.9e}) of the  evolution equation corresponds to the
resummation of QCD radiation from small angles 
(of the order of $M_0/p_{T}^{\, \gamma}$) up to large angles (of the order of
$M_F/p_{T}^{\, \gamma}$) with respect to the photon direction.

\vspace{0.3cm}
\noi
Owing to the physical picture that we have just described, we can easily 
evaluate
the $\ln R$ terms that are produced by partitioning the phase space through the
introduction of a cone of small size $R$.
We consider the fragmentation function $D(n, {\cal M} )$ where the scale 
${\cal M}$, which is of the order of $p_T^{\gamma}$ (the precise value of 
${\cal M}$ does not affect the following discussion at the level of the LL 
approximation), corresponds to the phase space region in which the collinear 
approximation neglecting the virtuality $p_a^2$ of parton $a$, cf. fig. 2, in 
the cross section of the  $2 \to 2$ subprocess holds. 
The choice ${\cal M} \sim p_T^{\gamma}$ implies that $D(n, {\cal M} )$ embodies 
collinear radiation up to very large angles with the respect to the photon 
direction. Therefore we can introduce the effect of the small-size cone by 
splitting $D(n, {\cal M} )$ in two parts (see eqn.~(\ref{2.15ber})).
To perform the splitting, we consider the expression (\ref{2.9e}) with 
$M_F={\cal M}$ and we rewrite its right-hand side as follows:
\bea
\label{2.13e}
D(n, {\cal M} ) 
& = & 
\;\; 
\int_{M_0^2}^{(Rp_{T}^{\gamma})^2} {dk^2 \over k^2}\ K(n)\ 
e^{\int_{k^2}^{(R p_{T}^{\gamma})^2} {dk{'}^2 \over k{'}^2} P(n)} \cdot 
e^{\int_{(R p_{T}^{\gamma})^2}^{{\cal M}^2} {dk{'}^2 \over k{'}^2} P(n)} \nn \\
&&
+ \int_{(Rp_T^{\gamma})^2}^{{\cal M}^2} {dk^2 \over k^2}\ K(n) \ 
e^{\int_{k^2}^{{\cal M}^2}  {dk{'}^2 \over k{'}^2} P(n)} \nn \\
&&
+ \;e^{\int_{(R p_{T}^{\gamma})^2}^{{\cal M}^2} {dk{'}^2 \over k{'}^2} P(n)}
\cdot e^{\int_{M_0^2}^{(R p_{T}^{\gamma})^2} {dk{'}^2 \over k{'}^2} P(n)}
\;D(n, M_0 )\;\;.
\eea

\noi 
In the LL approximation the splitting on the right-hand side
of eqn.~(\ref{2.13e}) acquires a geometrical 
meaning. We denote this approximation by introducing superscripts ``$^{(0)}$" 
and we rewrite eqn.~(\ref{2.13e}) as follows:
\beq
\label{2.15ber}
D^{(0)}(n,{\cal M}) 
= 
E^{(0)}(n, {\cal M},Rp_T ^\gamma) \ D^{(0)}(n,Rp_T ^\gamma)+
D^{(0)}(n,{\cal M},Rp_T ^\gamma) \;\;.
\eeq
The first term 
on the right-hand side of eqn.~(\ref{2.15ber}) corresponds to the 
configurations
in which there are partons inside the cone of radius $R$. 
This term has two factors.
The factor $D^{(0)}(n,Rp_T ^\gamma)$ 
is the fragmentation function
at the factorization scale $Rp_T ^\gamma$ (see eqn.~(\ref{2.9e})),  
\bea
\label{2.14e}
D^{(0)}(n,Rp_T ^\gamma) 
& = &
\int_{M_0^2}^{(Rp_T^\gamma )^2} {dk^2 \over k^2} \ 
K^{(0)}(n) \
e^{\int_{k^2}^{(Rp_T^\gamma )^2} \frac{dk^{\prime \, 2}}{k^{\prime \, 2}} 
\frac{\alpha_{s}(k^{\prime \, 2})}{2 \pi} \, P^{(0)}(n)} \nn \\
&&
+ \;\; e^{\int_{M_0^2}^{(Rp_T^\gamma )^2} \frac{dk^{\prime \, 2}}{k^{\prime \, 2}} 
\frac{\alpha_{s}(k^{\prime \, 2})}{2 \pi} \, P^{(0)}(n)} 
\;D^{(0)}(n,M_0) \;\;,
\eea
\noi 
and it corresponds to the contribution of all emitted partons that
are contained in the cone of radius $R$. The accompanying factor
$E^{(0)}(n, {\cal M}, R p_T^\gamma)$ is the following exponential factor
\begin{equation}\label{def-E}
E^{(0)}(n, {\cal M}, R p_T^\gamma) = 
e^{\int_{(Rp_T^\gamma )^2}^{{\cal M}^2} \frac{dk^{\prime \, 2}}{k^{\prime \, 2}}
\frac{\alpha_{s}(k^{\prime \, 2})}{2 \pi} \, P^{(0)}(n)} \;,
\end{equation}
and it sums the effect of the partons emitted outside the cone. 
The second term on the right-hand side of eqn.~(\ref{2.15ber})
is
\begin{equation}\label{remainder}
D^{(0)}(n, {\cal M}, R \, p_{T}^{\gamma} )
= 
\int_{(Rp_T^{\gamma})^2}^{{\cal M}^2} \frac{dk^2}{k^2} \ K^{(0)}(n) \ 
e^{\int_{k^2}^{{\cal M}^2} \frac{dk^{\prime \, 2}}{k^{\prime \, 2}}
\frac{\alpha_{s}(k^{\prime \, 2})}{2 \pi} \,  P^{(0)}(n)} \;,
\end{equation}
and it corresponds to the configurations with no partons inside the cone
of radius $R$. 

\vspace{0.3cm}
\noi
We then consider eqn.~(\ref{2.10e-new}) with $M_F={\cal M}$.
Since ${\cal M} \sim p_T^\gamma$, the higher-order contributions 
$\sigma^{(k)}_{F}$ and $\sigma^{(k)}_{D}$ 
(see eqns.~(\ref{2.11e}) and (\ref{2.11edir})) to $\sigma_{F}(n,{\cal M})$
and $\sigma_{D}(n,{\cal M})$ cannot produce a LL dependence
on $\ln R$ (roughly speaking, these higher-order contributions are due to parton
radiation at very large angles with respect to the photon). Therefore, we can
replace $\sigma_{F}(n,{\cal M})$
and $\sigma_{D}(n,{\cal M})$ with their Born level contribution and,
inserting eqn.~(\ref{2.15ber}) in eqn.~(\ref{2.10e-new}) 
(with $M_F={\cal M}$), we finally
obtain the expression of the cross section that explicitly shows the structure of
all the LL terms produced by the introduction of the auxiliary
cone of radius $R$. This final expression is
\bea
\label{2.19enew}
\sigma(n) 
& = & {\alpha_s \over \pi} \;\sigma^{Born}_{D}(n) \,
 + \, \left ( {\alpha_s \over \pi} \right )^2 \sigma_{F}^{Born}(n) 
 \; D^{(0)}(n, {\cal M}, Rp_T^\gamma) \nn \\
&&
+ \,
\left ( {\alpha_s \over \pi} \right )^2 \sigma_{F}^{Born}(n)
 \; E^{(0)}(n, {\cal M}, Rp_T^\gamma) \; D^{(0)}(n,Rp_T^\gamma) + \;\;\dots \;,
\eea
where the dots on the right-hand side denote contributions beyond the leading
logarithmic approximation.

\vspace{0.3cm}
\noi
The first-order perturbative expansion of the expression (\ref{2.19enew})
can directly be compared with the discussion of sec.~\ref{nlo-calc-R}.
Expanding eqs.~(\ref{2.14e})--(\ref{remainder}) up to the first order,
we obtain
\beq
\label{2.15enew}
D^{(0)}(n, R p_T ^\gamma) 
= D^{(0)}(n, M_0) \,+ \,
K^{(0)}(n) \ \ln \left( {(R p_T ^\gamma)^2 \over M_0^2} \right) + 
\frac{\alpha_{s}}{2 \pi} \, P^{(0)}(n)
\ \ln \left( {(R p_T ^\gamma)^2 \over M_0^2} \right) D^{(0)}(n, M_0)
+ \; \cdots \;\;,
\eeq
\beq
\label{e0exp}
E^{(0)}(n, {\cal M}, R p_T^\gamma) = 1 + 
\frac{\alpha_{s}}{2 \pi} \, P^{(0)}(n) \,
\ln \left( {{\cal M}^2 \over (Rp_T^\gamma )^2}\right) 
+ \; \cdots \;\;,
\eeq
\beq
\label{d0exp}
D^{(0)}(n, {\cal M}, R p_T ^\gamma) = 
K^{(0)}(n) \, \ln \left( {{\cal M}^2 \over (Rp_T^\gamma )^2}\right)
+ \; \cdots \;\;.
\eeq
Note that the contribution from single-parton radiation inside the cone depends
logarithmically on the ratio $(R p_T ^\gamma)^2/M_0^2$ (see
eqn.~(\ref{2.15enew})),
while the analogous contribution from ouside the cone depends
logarithmically on the ratio ${\cal M}^2/(Rp_T^\gamma )^2$ 
(see eqns.~(\ref{e0exp}) and (\ref{d0exp})).
Note also that, in eqns.~(\ref{2.15enew})--(\ref{d0exp}), the terms that are
proportional
to $K^{(0)}$ derive from the HO correction to the Born level direct cross section
(see fig.~\ref{fig1}), while those that are proportional to
$P^{(0)}$ derive from the HO correction to the Born level fragmentation
cross section (see fig.~\ref{fig2}). Inserting eqns.~(\ref{2.15enew})--(\ref{d0exp})
in the expression (\ref{2.19enew}), we obtain the HO contributions to the 
Born level cross sections. The HO contribution to the
Born level direct cross section is
\beq
\left ( {\alpha_s \over \pi} \right )^2 \sigma_{F}^{Born}(n)
\left[
K^{(0)}(n) \ \ln \left( {(R p_T ^\gamma)^2 \over M_0^2} \right) + 
K^{(0)}(n) \, \ln \left( {{\cal M}^2 \over (Rp_T^\gamma )^2}\right)
\right] \;\;,
\eeq
\vspace{0.3cm}
%
\noi 
where the first term reproduces the result of eqn.~(\ref{2.1e})
(regularized by $M_0^2$) and the second term reproduces 
the logarithmic dependence on
$R$ coming from the phase space part outside the cone
(see sec.~\ref{nlo-calc-R}).
Analogously, the HO contribution to the
Born level fragmentation component (fig.~\ref{fig2}) is
\beq
\left ( {\alpha_s \over \pi} \right )^2 \sigma_{F}^{Born}(n) \;
{\alpha_s \over 2\pi}
\left[
P^{(0)}(n) \ \ln \left( {(R p_T ^\gamma)^2 \over M_0^2} \right) + 
P^{(0)}(n) \, \ln \left( {{\cal M}^2 \over (Rp_T^\gamma )^2}\right)
\right] D^{(0)}(n, M_0) \;\;,
\eeq
where the two terms reproduce the logarithmic contributions discussed in the 
final part of sec.~\ref{nlo-calc-R}.

\vspace{0.3cm}

\noindent
Until now, we have explicitly considered the inclusive cross
section and no isolation criterion has been imposed 
on the partonic (hadronic) transverse energy inside the small-size cone.
Therefore, 
the LL dependence on $R$ actually cancels on the
right-hand side of eqn.~(\ref{2.19enew}) 
(the addition of the last two terms in the right-hand side  
indeed reconstructs the
fragmentation function $D^{(0)}(n,{\cal M})$, which is independent of $R$
order-by-order in the perturbative expansion). The discussion of the present
section aimed at paving the way for the isolated case. Indeed, the resummation
of the $\ln R$ dependence in the isolated cross section can straightforwardly be
carried out on the basis of the decomposition in the right-hand side
of eqn.~(\ref{2.19enew}).

\mysection{Isolated cross section and resummation 
of the $\ln R$ dependence}\label{isol}

To study the isolated cross section we use a more refined notation.
We explicitly reintroduce the parton indices, and  
we return (from Mellin space) to the 
configuration space, since the isolation constraint on transverse energies
is directly applied to momentum fractions
(see, e.g., the constraint on $z$ in eqns.~(\ref{2.4e}) and (\ref{2.5e})).

\vspace{0.3cm} 
\noi
The isolated cross section 
$d\sigma^{{is}}/dp_T^{\gamma}d\eta^{\gamma}$ with the cone isolation criterion of
eqns.~(\ref{e1-1}) and (\ref{e1-2})
is simply denoted by $\sigma^{{is}}(p^{\gamma};z_{cut},R)$, with 
(see eqn.~(\ref{2.4e}))
\begin{equation}
\label{zcutR}
z_{cut}  
= \frac{p_T^{\gamma}}{E_T^{iso} + p_T^{\gamma} } \;\;.
\end{equation}
The corresponding QCD factorization formula (analogous to eqn.~(\ref{2.10e-new}))
is written as in eqn.~(4.14) of ref.~\cite{1r}:
\begin{eqnarray}
\label{isxsR}
\sigma^{{is}}(p^{\gamma};z_{cut},R)
\!\!\!\!\!&=&\!\!\!\!\! \sum_{a} \int_0^1 \frac{dz}{z} \;
\widehat{\sigma}^{a,{is}}\left(\frac{p^{\gamma}}{z};
\frac{z_{cut}}{z},R;\mu,M,M_{F}\right) 
D_{a}^{\gamma}(z;M_F) \;\Theta(z-z_{cut})  \\
\!\!\!\!\!&+&\!\! 
\widehat{\sigma}^{\gamma,{is}}(p^{\gamma};z_{cut},R;\mu,M,M_{F}) \;.\nonumber
\end{eqnarray}
The subprocess cross sections $\widehat{\sigma}^{a,{is}}$ and 
$\widehat{\sigma}^{\gamma,{is}}$ are obtained by convolutions of 
partonic cross sections with the parton distribution functions of the two
colliding hadrons. These convolutions are not explicitly denoted throughout the
paper. The scale $M$ is the factorizaion scale of the parton distribution 
functions, and $\mu$ is the renormalization scale of the QCD coupling 
$\alpha_s$.

\vspace{0.3cm} 
\noi
The QCD perturbative expansion of the partonic cross sections 
leads to a corresponding expansion of the subprocess cross sections.
We write the expansion as follows (see eqns.~(4.18) and
(4.19) in ref.~\cite{1r}):
\begin{eqnarray}
\!\!\!\!\!\!\!\!\!\!\!\!
\widehat{\sigma}^{\gamma,{is}}(p;z_c,R;\mu ,M,M_{F}) \!\!\!\!\!&=&\!\!\!\!\! 
\left ( \frac{\alpha_s(\mu )}{\pi} \right ) \!
\sigma^{Born}_{\gamma}(p;M) + 
\left( \frac{\alpha_s(\mu)}{\pi} \right)^{\!2} \!\!
\sigma_{HO}^{\gamma,{is}}(p;z_c,R;\mu,M, M_{F}) + {\cal O}(\alpha_s^3) \,, 
\label{isaR} \\
\!\!\!\!\!\!\!\!\!\!\!\!
\widehat{\sigma}^{a,{is}}(p;z_c,R;\mu,M,M_{F}) \!\!\!\!\!&=&\!\!\!\!\! 
\left( \frac{\alpha_s(\mu)}{\pi} \right)^{\!2} \!\!
\sigma_{a}^{Born}(p;M) + 
\left( \frac{\alpha_s(\mu)}{\pi} \right)^{\!3} \!\!
\sigma^{a,{is}}_{HO}(p;z_c,R;\mu,M,M_F) +\! {\cal O}(\alpha_s^4) \,. 
\label{isbR}
\end{eqnarray}

\vspace{0.3cm} 
\noi
The evaluation of the isolated cross section $\sigma^{{is}}$ up to the NLO 
requires the computation of the first two terms on the right-hand side of 
eqns.~(\ref{isaR}) and (\ref{isbR}). This NLO computation, with the exact
dependence on $R$ (i.e., without any small-$R$ approximations) is performed in 
ref.~\cite{1r} and it is implemented in the programme {\tt Jetphox}.

\vspace{0.3cm} 
\noi
As discussed in the previous sections, at small values of $R$, the HO terms
$\sigma_{HO}^{\gamma,{is}}$ and $\sigma^{a,{is}}_{HO}$ in eqns.~(\ref{isaR}) and
(\ref{isbR}) contain a contribution that is proportional to $\ln R$. Additional
powers of $\ln R$ appear at still higher orders in the $\alpha_s$ expansion.
The direct component $\widehat{\sigma}^{\gamma,{is}}$ in eqn.~(\ref{isaR})
contains logarithmic
terms of the type $\alpha_s^{m+1} (\alpha_s \ln R)^k$, 
and the fragmentation component $\widehat{\sigma}^{a,{is}}$
in eqn.~(\ref{isbR}) contains logarithmic
terms of the type $\alpha_s^{m+2} (\alpha_s \ln R)^k$.
The LL terms are those with $m=0$ (and $k=1,2,3,\dots$).

\vspace{0.3cm} 
\noi
The resummation of the LL terms 
(the subscript notation $[ \;\;\; ]_{\rm LL}$ denotes the LL accuracy)
leads to the following result
\begin{eqnarray}
\label{isllR}
&&\!\!\!\!\!\!\!\!\!\!\!\!
\left[ \sigma^{{is}}(p^{\gamma};z_{cut},R) \right]_{\rm LL} 
=
\frac{\alpha_s(\mu )}{\pi} \;
\sigma^{Born}_{\gamma}(p^\gamma;M) 
+ \left( \frac{\alpha_s(\mu)}{\pi} \right)^{2}
\sum_{a} \int_0^1 \frac{dz}{z} \;
\sigma_{a}^{Born}\!\!\left(\frac{p^{\gamma}}{z};M\right) 
D_{a}^{(0)}(z;{\cal M},R\, p_T^{\gamma})  
\nonumber \\
&&\!\!\!\!\!\!\!\! +
\left( \frac{\alpha_s(\mu)}{\pi} \right)^{2}
\sum_{a, b} \int_0^1 \frac{dz}{z} \;
\sigma_{a}^{Born}\!\!\left(\frac{p^{\gamma}}{z};M\right) 
\int_z^1 \frac{dx}{x} 
\;E^{(0)}_{a b}\!\!\left(\frac{z}{x};{\cal M},R\, p_T^{\gamma} \right)
D_{b}^{\gamma (0)}(x;R\, p_T^{\gamma}) \;\Theta(x-z_{cut}) \;,
\end{eqnarray}
where $\sigma^{Born}_{\gamma}$ and $\sigma_{a}^{Born}$ are the Born level
subprocess cross sections in eqns.~(\ref{isaR}) and (\ref{isbR}), and
$E^{(0)}_{a b}$ and $D_{a}^{(0)}$ are the 
customary QCD evolution operators
at the LL order.
The expression of 
the parton evolution operator 
$E^{(0)}_{a b}(z;{\cal M},R\, p_T^{\gamma})$ is obtained 
from eqn.~(\ref{def-E}) by reinserting the
explicit dependence on the parton indices; the $n$-moments with respect to 
the momentum fraction $z$ are given by the exponentiated formula
(\ref{def-E}) by replacing the flavour NS kernel $P^{(0)}(n)$ with the DGLAP
matrix kernel $P^{(0)}_{ab}(n)$. A corresponding replacement is applied to
eqn.~(\ref{remainder}) to obtain the photon evolution operator
$D_{a}^{(0)}(z;{\cal M},R\, p_T^{\gamma})$; the explicit expression of this
operator is
\begin{eqnarray}
\label{daR}
D_{a}^{(0)}(z;{\cal M},R\, p_T^{\gamma}) = 
\sum_{b} \int_{(Rp_T^{\gamma})^2}^{{\cal M}^2} \frac{dk^2}{k^2} \;
\int_z^1 \frac{dx}{x} \;
\;E^{(0)}_{a b}\!\!\left(\frac{z}{x};{\cal M},k\right)
K_b^{(0)}(x) \;\;.
\end{eqnarray}

\vspace{0.3cm}
\noindent
The LL resummation formula (\ref{isllR}) is directly derived by supplementing
the right-hand side of eqn.~(\ref{2.19enew}) with the isolation constraint on 
the partonic transverse energy. Each of the three terms in 
the right-hand side of eqn.~(\ref{isllR}) is in direct correspondence with the
analogous term in eqn.~(\ref{2.19enew}). The first term in 
the right-hand side of eqn.~(\ref{2.19enew}) is the Born level direct
contribution, and the second term corresponds to kinematical configurations with
no partons inside the isolation cone. Since in these two cases the photon is
evidently isolated, these terms give the same contribution to the inclusive cross
section of eqn.~(\ref{2.19enew}) and to the isolated cross section of 
eqn.~(\ref{isllR}). In the third term on the right-hand side 
of eqn.~(\ref{2.19enew}), the operator $E^{(0)}$ embodies parton radiation
outside the isolation cone, while the photon fragmentation function 
$D^{(0)}(n, R\, p_T^{\gamma})$ embodies parton radiation inside the cone.
Therefore, only
the fragmentation function must be isolated by applying the transverse-energy
isolation constraint of eqn.~(\ref{e1-2}).
This isolation constraint leads to the momentum fraction cut $x > z_{cut}$ that
is explicitly implemented in eqn.~(\ref{isllR}).

\vspace{0.3cm}
\noindent
In the LL resummed expression (\ref{isllR}), the fragmentation function
$D_{b}^{\gamma (0)}$ is evaluated at the evolution scale $Rp_T^\gamma$.
Therefore, it is interesting to make a comparison of eqn.~(\ref{isllR})
with the NLO cross section by choosing $M_F=Rp_T^\gamma$ in the NLO expression.
Setting $M_F=Rp_T^\gamma$ in eqns.~(\ref{isxsR})--(\ref{isbR}), the NLO result
effectively resums the $\ln R$ terms produced by parton radiation inside the
isolation cone. However, the corresponding HO subprocess cross sections
$\sigma_{HO}^{\gamma,{is}}(M_F=Rp_T^\gamma)$ and 
$\sigma^{a,{is}}_{HO}(M_F=Rp_T^\gamma)$ of eqns.~(\ref{isaR}) and
(\ref{isbR}) still contain a residual $\ln R$ term (which is due to 
partons radiated outside the isolation cone) that is not resummed by the NLO
result. In the resummation formula (\ref{isllR}), this residual 
$\ln R$ term at NLO is produced by the first-order expansion of
$D_{a}^{(0)}(z;{\cal M},R\, p_T^{\gamma})$ 
(which contributes to $\sigma_{HO}^{\gamma,{is}}(M_F=Rp_T^\gamma)$)
and $E^{(0)}_{a b}(z/x;{\cal M},R\, p_T^{\gamma})$
(which contributes to $\sigma^{a,{is}}_{HO}(M_F=Rp_T^\gamma)$).

\vspace{0.3cm}
\noindent
Using eqn.~(\ref{daR}) and the exponentiated form (see eqn.~(\ref{def-E}))
of the evolution operator 
$E^{(0)}({\cal M},R\, p_T^{\gamma})$, the expression (\ref{isllR})
can be used to explicitly resums the LL contributions to the isolated cross
section at small values of $R$. Note, however, that the scale
$Rp_T^\gamma$ must be sufficiently larger than the typical scale of the
non-perturbative domain
(e.g., $Rp_T^\gamma \geq M_0$ where $M_0$ is the cut-off scale in 
eqn.~(\ref{2.9e})). At very small values of $R$ (and $Rp_T^\gamma$), the photon
fragmentation function 
$D_{b}^{\gamma}(x;R\, p_T^{\gamma})$ in eqn.~(\ref{isllR}) and, more generally,
the isolated cross section become sensitive to sizeable non-perturbative effects
that are not taken into account by the
perturbative QCD factorization formula (\ref{isxsR}).

\vspace{0.3cm}
\noi The resummation of the $\ln R$ terms can be generalized beyond the LL level of
eqn.~(\ref{isllR}). We have worked out the formal generalization to arbitrary
logarithmic accuracy. The all-order generalization and the explicit treatment
of next-to-leading logarithmic (NLL) terms will be presented in a forthcoming
paper. The extension of resummation to other isolation criteria
(e.g., the criterion in sec.~\ref{impli-hgc}) is in progress. 
In the next section,
we explicitly apply the LL resummation formula of eqn.~(\ref{isllR}), and we
present ensuing quantitative results for Tevatron and LHC kinematical
configurations.

\mysection{Quantitative results}\label{results}

We have implemented the LL resummation of the $\ln R$ terms in the programme 
{\tt Jetphox}. This implementation supplements the complete NLO result
\cite{1r} (which has the exact dependence on $R$) with the resummation of all the
LL terms beyond the NLO. The complete NLO result is added to 
a `subtracted version' of the LL resummation formula (\ref{isllR}).
This subtracted version avoids double counting of perturbative terms.
It is obtained by considering the LL formula (\ref{isllR})
and by explicitly subtracting from it the
terms that are obtained by expanding the same formula up to the NLO.

\vspace{0.3cm}
\noindent
We add some comments about our actual implementation of resummation in 
{\tt Jetphox}. We want results with
consistent NLO accuracy (and exact dependence on $R$) for both 
the direct and fragmentation contributions and, therefore, we have to use
NLO fragmentation
functions (and parton distribution functions). To this purpose,
in the LL resummation formula (\ref{isllR}), 
the LO fragmentation function $D^{(0)}$ can be replaced by the NLO one: this
replacement is permitted, since it produces corrections beyond the LL
approximation.
Then we note that, 
due to the isolation cut $x > z_{cut}$,
the contribution of the fragmentation function $D_b^{\gamma}(x; Rp_T^\gamma)$
is small, so that the impact of the resummed 
factor $E^{(0)}({\cal M}, Rp_T^\gamma,)$ in the third term on the right-hand side of 
eqn.~(\ref{isllR}) is not significant.  
For the sake of numerical simplicity, this resummed factor is replaced by its truncation
at ${\cal O}(\alpha_{s})$:
\begin{equation}\label{trunc}
E^{(0)}(z/x; {\cal M}, Rp_T^\gamma)  
\to  
\delta(1-z/x) + \frac{\alpha_s}{2 \pi} \, P^{(0)} (z/x) \,  
\ln \left ( \frac{{\cal M}^{2}}{(R p_T^{\gamma})^{2}} \right) \;.
\end{equation}
We observe that the replacement in eqn.~(\ref{trunc}) implies that the  third
term on the right-hand side of  eqn.~(\ref{isllR}) does not produce any LL terms
beyond the NLO, provided the NLO perturbative expansion is carried out at the
factorization scale  $M_F=R p_T^{\gamma}$ (this observation is equivalent to
that made in the final part of section~\ref{isol}, where we pointed out that the
NLO result with  $M_F=Rp_T^\gamma$ effectively resums the LL terms produced by
parton radiation inside the isolation cone). Therefore, 
we can consider the NLO result
of {\tt Jetphox} with $M_F=R p_T^{\gamma}$ and supplement it with the LL terms
produced by the sole direct component of the cross section in eqn.~(\ref{isllR}).
In practical terms, the resummation part to be added to the NLO cross section
$\sigma^{{is}}(p^{\gamma};z_{cut},R)$
of eqns.~(\ref{isxsR})--(\ref{isbR}) is the following contribution:
\begin{eqnarray}
\label{added}
+ \left( \frac{\alpha_s(\mu)}{\pi} \right)^{2}
\sum_{a} \int_0^1 \frac{dz}{z} \;
\sigma_{a}^{Born}\!\!\left(\frac{p^{\gamma}}{z};M\right) 
\left\{ D_{a}^{(0)}(z;{\cal M},R\, p_T^{\gamma})
- K_a^{(0)}(z) \;\ln \frac{{\cal M}^2}{(Rp_T^{\gamma})^2}
\right\}.
\end{eqnarray}
This contribution corresponds to the first two terms in the right-hand side of 
eqn.~(\ref{isllR}), after subtraction of their NLO expansion which is already
contained in the ${\cal O}(\alpha_s^2(\mu))$ term of the
expression~(\ref{isllR}) (the second term in the curly bracket of
eqn.~(\ref{added}) is exactly  the first-order expansion of 
$D_{a}^{(0)}(z;{\cal M},R\, p_T^{\gamma})$, which is the first term in the 
curly bracket).

\vspace{0.3cm}
\noindent
In this section we compare the outputs of {\tt Jetphox} obtained without 
and with resummation.

\vspace{0.3cm}
\noindent
We start our presentation with kinematics relevant to Tevatron experiments 
that we already studied in ref.~\cite{1r}. We consider proton--antiproton
collisions at the centre--of--mass energy $\sqrt{s} = 1.8$~TeV, and
we use $p_T^\gamma = 15$~GeV, $-0.9 \leq \eta
^\gamma \leq 0.9$  and  $\varepsilon = E_T^{iso}/p_T^\gamma = 0.1333$
(i.e., $z_{cut}=0.88235$ and $E_T^{iso}=2$~GeV). 
The calculations are done with $N_f = 5$ flavours
of massless quarks. The renormalization scale ($\mu$) and the 
factorization scale ($M$) of the parton distribution functions 
are set to be equal to
$p_T^\gamma /2$~. As for the factorization scale of the photon
fragmentation function, we study the cases with  $M_F = p_T^\gamma/2$
(conventional scale) and $M_F = R p_T^\gamma$ (``cone scale'').  We
use the CTEQ6M parton distribution functions \cite{3r}. The fragmentation
functions are those of the BFG 
set II \cite{2r}. As we have already discussed, 
we use NLO fragmentation
functions and parton distribution functions.
The resummed contribution 
of eqn.~(\ref{added}) is calculated with 
${\cal M} = p_T^\gamma$.

\vspace{0.3cm}

\noindent
Our results are summarized in tables~\ref{tab1} and \ref{tab2}. Table~\ref{tab1} gives the
details of the direct and fragmentation contributions to the isolated
cross sections calculated with the ``conventional''
and ``cone'' scales, with or without resummation. We also give the results of
the non-isolated cross
sections. In table 2 one can find the behavior with $R$ of the total cross sections. 
The numerical values 
that we obtain at the Born level and at the NLO 
are not identical to those of ref.~\cite{1r}, since a different set
(the MRST-99 set) of parton distribution functions was used 
in ref.~\cite{1r}.  \par

\begin{table}[htb]
\begin{center}
\hskip - 0.5 truecm \begin{tabular}{|c|c|c|c|c|c|c|c|c|}
\hline
\multicolumn{5}{|c|}{\bf DIRECT} &\multicolumn{4}{|c|}{\bf FRAGMENTATION} \\
\hline
${R}$ &\mbox{\small Born} &\mbox{\small NLO} &\mbox{\small NLO} &\mbox{\small NLO} &\mbox{\small Born} &\mbox{\small Born} &\mbox{\small NLO} &\mbox{\small NLO} \\
 & & $p_T^\gamma /2$ & $R p_T^\gamma$ & $R p_T^\gamma$ & $p_T^\gamma /2$ & $Rp_T^\gamma$ & $p_T^\gamma /2$ & $Rp_T^\gamma$ \\
&&&&\mbox{\footnotesize -resummed} &&&& \\
\hline
.9 &1972 &3737 &3586 &3592 &291 &348 &516 &700 \\
.7 &1972 &3951 &3864 &3851 &291 &324 &555 &662 \\
.5 &1972 &4197 &4197 &4171 &291 &291 &597 &597 \\
.3 &1972 &4532 &4663 &4593 &291 &245 &654 &496 \\
.1 &1972 &5203 &5616 &5294 &291 &165 &764 &318 \\
\hline
{\small No isol} &1972 &3655 & & &1875 & & 2044 & \\
\hline
\end{tabular}
\caption{Variation with $R$ of the various contributions to the cross sections
(the values are expressed in pb/GeV)
at $\sqrt{s} = 1.8$~TeV.}
\label{tab1}
\end{center}
\end{table}

\noindent
The first point to note in table~\ref{tab1} is the strong effect of isolation on the fragmentation
component. Another noticeable point is the increase of the NLO direct contribution when $R$
decreases; this increase is in agreement with the $\ln R$ enhancement shown in the
expression (\ref{2.5e}). In particular, the isolated contribution can
be larger than the non-isolated contribution (we report the non-isolated
reference values at
the factorization scale $M_F= p_T^\gamma /2$).
No resummation is
involved in the calculation of the non-isolated cross-section. 
The
resummed direct cross section is smaller (by about 5\% at $R = 0.1$) than the cross section
without resummation. 
The fragmentation component is quite sensitive to
$R$ already at the Born level through the $M_F$ dependence on $R$
(this is because the typical behaviour of the fragmentation function
is $D(Rp_T^\gamma) \sim \ln(Rp_T^\gamma/M_0)$). 

\begin{table}[htb]
\begin{center}
\begin{tabular}{|c|c|c|c|}
\hline
$R$ &     NLO         &      NLO       &    NLO \\
    & $p_T^\gamma /2$ & $R p_T^\gamma$ & $R p_T^\gamma$ \\
    &                 &                & \mbox{\footnotesize -resummed} \\
\hline
.9 &4253 &4286 &4292  \\
.7 &4506 &4526 &4513 \\
.5 &4794 &4794 &4768  \\
.3 &5186 &5159 &5089\\
.1 &5967 &5934 &5612 \\
\hline
{\small No isol
} &5699 & & \\
\hline
\end{tabular}
\caption{Variation with $R$ of the total cross sections (in pb/GeV)
at $\sqrt{s} = 1.8$~TeV.}
\label{tab2}
\end{center}
\end{table}

\noindent
The NLO cross sections are given in table~\ref{tab2}. In the first column, at the
value $R = 0.1$ we notice a violation of
unitarity since the non-isolated cross section at the reference scale
$M_F = p_T^{\gamma}/2$ is 
smaller than the isolated one for both the scale choices $M_F =p_T^{\gamma}/2$ 
and $M_F = Rp_T^\gamma$.
When $R$ becomes small, the $\ln R$ terms has to be resummed:
this is performed in the rightmost column of table~\ref{tab2}, and we notice that resummation 
does restore unitarity. 
Notwithstanding we also notice that at $R =0.1$
the resummed cross section is about 5\% smaller than 
the NLO cross section.
Therefore 
the effect of resummation is 
not very large at $R =0.1$, although it is qualitatively and
conceptually important regarding the restoration of 
unitarty. We cannot explore lower values of $R$ because the scale at which the fragmentation
function is calculated becomes too small.\par

\vspace{0.3cm}

\noindent
To study smaller values of $R$ we turn to the LHC kinematics and we consider
higher values of $p_T^\gamma$. We consider 
proton--proton collisions at $\sqrt{s} = 7$~TeV, $p_T^\gamma = 100$~GeV and $|\eta^\gamma
| < 0.6$. The energy isolation parameter is $\varepsilon = 0.04$
(i.e., $z_{cut}=0.9615$ and $E_T^{iso}=4$~GeV) and the radius of the isolation cone 
is varied in the range $0.06 \leq R \leq 0.5$. 
The parton distributions functions, the fragmentation functions and the scale
choices are the same as in the Tevatron study reported in tables~\ref{tab1} and 
\ref{tab2}. Our LHC results
are summarized in table~\ref{tab3}.\par

\vspace{0.3cm}

\noindent
The first two columns of table 3 show that the NLO cross section is very stable with respect to
changes of the factorization scale $M_F$ when the photon is isolated. Here we also note a violation
of unitarity 
for small
values of $R$ ($R \leq 0.1$). The rightmost column displays the effect of resummation. 
It is small, and it is of
the order of 7\% at $R = 0.06$. Unitarity is no more violated down to the very small value 
$R = 0.06$. This very small value of $R$ is however 
extreme; the experimental isolation
cones typically have a radius bigger than $0.3$ for which the effect of resummation is even smaller 
($\leq 1$\%).

\begin{table}[htb]
\begin{center}
\begin{tabular}{|c|c|c|c|}
\hline
$R$ &     NLO         &      NLO       &    NLO \\
    & $p_T^\gamma /2$ & $R p_T^\gamma$ & $R p_T^\gamma$ \\
    &                 &                & \mbox{\footnotesize -resummed} \\
\hline
.5 &3.59 &3.59 &3.57  \\
.3 &3.86 &3.85 &3.81 \\
.1 &4.35 &4.34 &4.19  \\
.06 &4.56 &4.55 &4.24 \\
\hline
{\small No isol
} &4.29 &  &  \\
\hline
\end{tabular}
\caption{Variation with $R$ of the total cross sections (in pb/GeV) at $\sqrt{s} = 7$~TeV.}
\label{tab3}
\end{center}
\end{table}

\vspace{0.3cm}

\noindent
The issue of the violation of unitarity in the NLO calculation of the isolated 
cross section with narrow cones is quantified, throughout this section,
by the numerical comparison with the non-isolated photon cross section
at the reference fragmentation scale $M_{F} = p_{T}^{\gamma}/2$.
Obviously the choice of reference scale has some degree of arbitrariness, 
and any other commonly used choice, such as
$M_F = p_{T}^{\gamma}$ or $2 \, p_{T}^{\gamma}$ would 
affect the numerical results somewhat (we have explicitly checked that the use
of these scales does not significantly change the values of the non-isolated
cross section).
However the main trends of the results presented in this section 
are evident, 
and are summarized as follows.  
1) The violation of unitarity observed in NLO cross sections for photons isolated with 
narrow cones (with aperture $R \ll 1$) is due to the inappropriate truncation, 
to fixed order, of an expansion that involves terms logarithmically enhanced in
$\ln R$. The violation cannot 
be cured by simply adjusting the scales.
Instead the resummation of these logarithmically enhanced 
terms, as we have performed, cures this unitarity puzzle down to values of $R$
that are 
much lower than the ones physically used in colliders experiments.
2) Notwithstanding, down to at least $R = 0.3$, which is already smaller 
than the value 0.4 commonly 
used in colliders experiments, this resummation amounts to a small, 
few \% correction with respect 
to the NLO calculation using a standard choice of $M_F$, 
and this correction is smaller than 
the usual scale uncertainty of the NLO prediction. The unitarity puzzle 
is therefore {\em not} an issue 
in NLO predictions with the cones sizes that are commonly used 
in colliders experiments.
Moreover, resummation can be regarded and used as a complementary and additional
theoretical tool that is available to asses the reliability of the QCD
predictions and to quantify their uncertainties in studies with moderately-small
values of $R$.

%% file: hollow-short.tex
\mysection{Implications for a hollow-cone criterion}\label{impli-hgc}

Using the standard cone criterion or an alternative one, such as the Frixione
criterion  \cite{frixione} or one of its discretized versions
\cite{frix-discretise}, it may be experimentaly  difficult to apply a cut on the
accompanying energy in the region where the electromagnetic shower  develops in
the detector, since it may be difficult to disentangle the accompanying energy
from the  photon energy inside that region. The shape and size of this region
are detector dependent. To first approximation, the electromagnetic shower
roughly fills a cone of radius $r \sim 0.1$:   considering the standard cone
criterion, this   would correspond to an inner narrow cone of inefficient
isolation inside the usual cone of radius $R$.  For a `hollow' cone with an
inner (empty) cone of such a small size ($r \sim 0.1$),  the issue of the
resummation of $\ln r$ contributions can matter. In this section, we perform a
first explorative investigation of this issue.

\vspace{0.3cm}
\noindent 
We consider a hollow-cone variant of the standard cone isolation
criterion in eqns.~(\ref{e1-1}) and (\ref{e1-2}). A cone of radius $R$ around
the photon direction is still considered, but the upper limit $E_{T}^{iso}$
is enforced on the hadronic transverse energy inside an annular region 
${\cal C}_{\gamma}^{holl.}(R,r)$ of width $R-r$,
rather than on the hadronic transverse energy inside the whole cone.
The constraint in eqn.~(\ref{e1-2}) is thus replaced by
\begin{equation}\label{holconst}
\sum_{k \,\in \,{\cal C}_{\gamma}^{holl.}(R,r)} E_{T}^{(k)} \, 
< E_{T}^{iso} \;\;.
\end{equation}
Therefore, {\em no} isolation is applied 
inside the innermost region delimited by the small cone of radius $r$ 
(thus coined `hollow', in the sense of free of any isolation
constraint, in what follows).
The usual isolation is instead implemented {\em outside} the innermost cone 
of radius $r$.
 
\vspace{0.3cm}
\noindent 
Note that the hollow-cone criterion selects photons that are {\em
less} isolated than those selected by the standard cone criterion. Therefore,
at fixed values of $E_{T}^{iso}$ for both criteria, the corresponding cross
sections fulfil the physical requirement
\begin{equation}\label{holbound}
\sigma(p^{\gamma}) \geq \sigma^{holl}(p^{\gamma};R,r)
\geq \sigma^{{is}}(p^{\gamma};R)  \;\;,
\end{equation}
where $\sigma(p^{\gamma})$ is the inclusive cross section with no isolation,
$\sigma^{{is}}(p^{\gamma};R)$ is the cone isolation cross section
(i.e., the cross section considered in sects.~\ref{isol} and \ref{results}))
and $\sigma^{holl}(p^{\gamma};R,r)$ is the cross section for 
the hollow-cone criterion.

\vspace{0.3cm}
\noindent 
The factorization structure of the perturbative QCD calculation of the 
hollow-cone
cross section is discussed in the final part of sec.~4 of ref.~\cite{1r}.
Here we briefly comment on the expected behaviour of the NLO cross section in
the case of a very small
innermost cone ($ r \to 0$).
As discussed in sec.~\ref{nlo-calc-R}, the presence of the boundary of a cone of
small radius $R_0$ (we use a generic $R_0$ to allow us to make a following
distinction between $R$ and $r$) around the photon direction produces 
$\ln R_0$-terms in the NLO calculation. Typically, single-parton radiation inside the
cone leads to {\em negative} $\ln R_0$-terms, while parton radiation just outside
the cone leads to analogous {\em positive} terms. Within the standard cone
criterion $(R_0=R)$, energy isolation is applied inside the cone thus
suppressing the negative $\ln R$-terms, and the NLO cross section receives
a total {\em positive} contribution from the $\ln R$-terms. 
In the hollow-cone case $(R_0=r)$, the situation is reversed with respect to the
standard cone criterion: energy isolation is now imposed outside the cone $r$ 
instead of inside it. This suppresses the positive $\ln r$-terms, and the 
NLO cross section $\sigma^{holl}(p^{\gamma};R,r)$
receives
a total {\em negative} contribution from the $\ln r$-terms. 
Eventually, in the limit $ r \to 0$ (at fixed $R$), we can expect a violation of
unitarity of the NLO result since $\sigma^{holl}(p^{\gamma};R,r)$ can become
smaller than $\sigma^{{is}}(p^{\gamma};R)$, thus violating the lower bound of
the physical requirement in eqn.~(\ref{holbound}).

\vspace{0.3cm}
\noi We use {\tt Jetphox} to compute the NLO cross section for the hollow-cone
criterion, and we present quantitative results by considering  
the LHC kinematical configuration already discussed in sec.~\ref{results}.
We consider  a hollow-cone criterion with the inner cone of radius $r= 0.1$ 
and the outer cone of radius $R = 0.4$. 
The results for the NLO cross sections are given in table~\ref{tab8}, where they
are also compared with the corresponding standard-cone cross section 
(with $R = 0.4$).

\vspace{0.3cm}
\noi 
The NLO hollow-cone cross section in table~\ref{tab8} is computed by using the
fragmentation-scale choice  $M_F = r \, p_T^\gamma$. This choice is motivated by
analogy with the discussion in the previous sections: setting 
$M_F = r \, p_T^\gamma$, we effectively resum part of the higher-order 
$\ln r$ terms (the part from parton radiation inside the inner cone). For the
sake of direct comparison, the NLO standard-cone cross section reported in 
table~\ref{tab8} is computed by using the same numerical value, 
$M_F = 0.1 \, p_T^\gamma$, that is used in the hollow-cone cross section.
The results of the standard-cone cross section that are obtained with different
choices of $M_F$ $(M_F =  p_T^\gamma/2 , \;M_F = R p_T^\gamma)$ at the NLO and
with LL resummation were reported in table~\ref{tab3}. All these values of the 
standard-cone cross section at $R = 0.4$ (the value in table~\ref{tab8}, and the
values that can be inferred from table~\ref{tab3}) are numerically very
similar. Therefore, in the context of the discussion in this section, we can
state that we have a reliable estimate of the standard-cone cross 
section at $R = 0.4$. \par

\vspace{0.3cm}
\begin{table}[here]
\begin{center}
\begin{tabular}{|c|cc|cc|c|} 
\hline 
\hbox{\bf cone}&\multicolumn{2}{|c|}{\bf direct}&\multicolumn{2}{|c|}{\bf fragmentation}&{\bf total}\\ 
\cline{2-6}
\hbox{\bf type} &{\bf Born} & {\bf NLO} & {\bf Born} & {\bf NLO} & {\bf NLO} \\ 
\hline
standard & 2.08 & 3.56 & .077 & .165 & 3.73 \\ 
hollow   & 2.08 & 3.09 & .91  & .43 &  3.52 \\ 
\hline 
\end{tabular}
\caption{NLO cross sections (in pb/GeV) at ${\sqrt s}=7$~TeV, with $\varepsilon
= 0.04$ and $M_F = r \, p_T^\gamma$.
Comparison of the isolated cross sections for the standard cone ($R = 0.4$) 
vs. the hollow-cone ($R =0.4, \,r=0.1$) crireria.}
\label{tab8}
\end{center}
\end{table}

\noi The results in table~\ref{tab8} show that the 
total (i.e. direct $+$ fragmentation) NLO predictions
for the standard cone criterion 
and the hollow-cone criterion do not differ much, despite the drastic absence 
of isolation in the inner cone. The two NLO results depart from each other 
by about $6$\%, and they 
would differ even less if some loose isolation were implemented inside 
the inner cone. Indeed, our hollow-cone criterion with {\em no} isolation 
inside the inner cone is an extreme simplification for modelling 
the experimental region of inefficient isolation (for instance, CMS analyses
impose a veto on charged tracks in the close vicinity of the
photon candidates).

\vspace{0.3cm}
\noi
Nonetheless, the quantitative results in table~\ref{tab8} also show that the
hollow-cone cross section is smaller than the standard cone cross section, thus
violating the bound $\sigma^{holl}(p^{\gamma};R,r)
\geq \sigma^{{is}}(p^{\gamma};R)$ that is set by the physical requirement in
eqn.~(\ref{holbound}). This violation is an artifact of the NLO truncation
of the perturbative QCD calculation. The misbehaviour of the NLO cross section
for the hollow-cone criterion is due to the presence of large and negative 
$\ln r$ terms, as discussed at the beginning of this section.
We have quantitatively checked that this misbehaviour of the NLO calculation 
persists and it is enhanced by decreasing further the values of $r$. In
particular, at very small values of $r$, the NLO corrections are found to be
very large especially for the fragmentation component of the cross section,
which eventually becomes (increasingly) negative.
The different relative importance of the fragmentation component 
compared with the direct component 
comes from the absence of isolation inside the inner cone.
Without the resummation of $\ln r$ terms, we cannot expect
a sound theoretical result for the hollow-cone criterion
at very small values of $r$. We postpone the study of resummation 
of $\ln r$ terms for
the hollow-cone criterion to future work, which is in progress. In particular,
in the hollow-cone case, very small values of $r$ and, consequently, very large
values of $\ln (1/r)$ are considered. A reliable quantitative treatment of
resummation effects may thus requires the LL contributions and the inclusion of
terms at the NLL level, which may have a not negligible quantitative role.

\vspace{0.3cm}

\mysection{Conclusions}\label{secsum}

We have considered the standard cone isolation criterion that is used to measure
prompt-photon cross sections at TeV colliders. We have performed a detailed
study of the dependence of the QCD theoretical  predictions on size $R$ of the
isolation cone. The dependence arises from the  mismatch of parton radiation in
the region inside the  cone, which is submitted to isolation, vs. the region
outside the cone,  where no isolation is imposed.  The dependence on the cone
size $R$ is dominantly logarithmic at small $R$,  and the NLO predictions 
become unreliable when $R$ becomes too small. We restored the reliability of 
the theoretical estimates by performing the resummation of this logarithmic
dependence to LL accuracy in $R$. The resummation eventually
amounts to the following procedure: the fragmentation scale is set to the value
$M_F \sim R \, p_{T}^{\gamma}$ (as simply suggested from physical insight) to
take into account the $\ln R$ dependence from parton radiation inside the cone,
and an additional explicit resummation is performed to control the $\ln R$
dependence that arises from  the region outside the cone, which is not submitted
to isolation. We have implemented this resummation in the partonic Monte Carlo
programme {\tt Jetphox}, which also includes the non logarithmic $R$ dependence
that is not negligible at the moderate value $R \sim 0.4$ that is used
experimentally at the LHC.  We have presented ensuing quantitative results that
show how the resummation cures the instabilities of the NLO calculation down to
very small values of  $R$. At the typical values of $R$ ($R \,\gsim \,0.3$) that
are currently used in collider experiments, the resummation effects are small,
and they are not larger than the size of the usual theoretical uncertainties of
the NLO predictions.

\vspace{0.3cm}
\noi 
We also explored the case of ``hollow cone" isolation.
The isolation is imposed in an annulus between two cones 
of radii $r$ and $R$, with $r < R$, whereas the regions outside the annulus, including the one
inside the inner cone, is free from any isolation constraints. The cone with
small radius $r$ aims at simulating the size of the electromagnetic shower that
develops in the calorimeter. Considering the values $r = 0.1$ and $R = 0.4$,
the corresponding NLO cross section 
does not differ much from the cross section with the standard cone isolation, 
and the difference would 
be even smaller by using some loose isolation inside the inner cone, rather 
than none as we did.  
However, we also noticed that the strict implementation of the hollow-cone
criterion (with the extreme situation of absolutely no isolation inside
the inner cone of radius $r$) can produce NLO inconsistencies, leading
to quantitative values of the hollow-cone cross section that are smaller 
than those of the 
more-isolated cross section of the standard cone isolation.
A proper resummation of $\ln r$ terms would be necessary to cure these 
NLO inconsistencies and to safely use the hollow-cone criterion
for perturbative QCD calculations at small values of $r$.

\vspace{0.5cm}
\noi {\bf Acknowledgements}. This work is supported in part by 
the Research Executive Agency (REA) of the European Union under 
the Grant Agreement number PITN-GA-2010-264564 (LHCPhenoNet).